\documentclass[aps,prl,twocolumn,floatfix,showpacs]{revtex4}

\usepackage{graphics}
\usepackage{graphicx}
\usepackage{amssymb}
\usepackage{amsfonts}
\usepackage{amsmath}
\usepackage{bm} 

\newcommand{\nab}{\bm{\nabla}}
\begin{document}

\title{Vortex Lattice Structural Transitions: 
a Ginzburg-Landau Model Approach}
\author{A. D. Klironomos}
\author{Alan T. Dorsey}
\affiliation{Department of Physics, University of Florida, P.O. Box 118440, Gainesville, Florida 32611-8440}
\date{\today}

\begin{abstract}

We analyze the rhombic to square vortex lattice phase transition in 
anisotropic superconductors using a variant of Ginzburg-Landau (GL) theory.  
The mean-field phase diagram is determined to second order in the 
anisotropy parameter, and shows a reorientation transition of the 
square vortex lattice with respect to the crystal lattice. We then 
derive the long-wavelength elastic moduli of the lattices, and use them to
show that thermal fluctuations produce a \textit{reentrant} 
rhombic to square lattice transition line, similar to recent studies 
which used a nonlocal London model. 

\end{abstract}


\maketitle

Vortex lattices in superconductors with four-fold anisotropy can exhibit a 
rhombic to square structural phase transition  
due to a competition between the anisotropy (favoring a square lattice) 
and the repulsive vortex-vortex interaction (favoring a triangular lattice).  
This transition has been observed in the 
borocarbide family of superconductors $R$Ni$_{2}$B$_{2}$C (with $R$ a rare 
earth element), whose anisotropy is 
due to lattice effects \cite{eskildsen,eskildsen2,exp3,sakata,vinnikov2},   
and more recently in the overdoped cuprate superconductor 
La$_{1.83}$Sr$_{0.17}$CuO$_{4+\delta}$ \cite{gilardi}, whose anisotropy is 
due to unconventional superconducting order.  
Recent small angle neutron scattering (SANS) data for 
LuNi$_{2}$B$_{2}$C \cite{eskildsen} 
show the transition line in the $H-T$ plane curving upward and avoiding 
$H_{c2}$, contrasting with mean-field predictions and underscoring the 
importance of thermal fluctuations even in this low temperature superconductor.
Some features of the data can be understood using a nonlocal London model 
developed by Kogan and collaborators \cite{kogan1,kogan2}; however, the 
London model is ostensibly only applicable at low magnetic inductions, leaving
as an open question the behavior close to $H_{c2}$. An alternate explanation
for the reentrant behavior of the vortex lattice transition is also available in
the literature \cite{nakai}.
 
In this Letter we study the effect of thermal fluctuations on the 
rhombic to square vortex lattice transition using 
an anisotropic GL model.  We first determine the mean-field phase
diagram to second order in the anisotropy parameter, and find that even 
at this level GL theory predicts a novel 
reorientation transition of the square vortex lattice with respect to the 
crystal lattice, an effect overlooked in previous GL studies. 
We then derive the (nondispersive) elastic 
moduli of the lattices using methods first pioneered by 
Brandt \cite{brandt1} for isotropic superconductors.  Finally, these moduli 
are used to determine the fluctuation renormalized phase boundary in the 
$H-T$ plane. For parameters typical of the borocarbides we find a 
reentrant phase boundary, in agreement with the nonlocal London model 
results of Ref.~\cite{gurevich}.
As GL theory is expected to be valid near $H_{c2}$, our 
results nicely complement and extend the nonlocal London model results.

Accounting for anisotropy, the linearized first GL equation (in dimensionless units) is
\begin{equation}
\label{GL}
\Pi^{2}\psi-\psi-\epsilon(\Pi_{x}^{2}-\Pi_{y}^{2})^{2}\psi=0,
\end{equation}
where ${\bf\Pi}=(\nab/i\kappa-{\bf A})$ is the covariant derivative and 
$\epsilon$ is the dimensionless anisotropy parameter.
For the cuprates the anisotropic term can be derived rigorously from the 
two field $s$ and $d$-wave GL free energy \cite{franz,heeb} by 
integrating out the $s$-wave term \cite{affleck}. For the conventional, 
but highly anisotropic, borocarbides it is a phenomenological term that 
will be sufficient to capture all the interesting physics.  
The symmetry breaking properties of the anisotropic term dictate the 
introduction of the orientation angle 
$\phi$ of the vortex lattice relative to the underlying crystal 
lattice---this term provides an effective coupling of the vortex and 
crystal lattices, and makes structural transitions possible.

We have generalized the virial theorem of superconductivity \cite{doria,klein}
to include the anisotropic term in the extended GL
theory \cite{future}, with the result 
\begin{eqnarray}
\label{virial}
\frac{1}{4\pi} {\bf H}\cdot{\bf B} & = & 
\bigg\langle|\psi|^{2}-|\psi|^4+\frac{{\bf B}^{2}({\bf
x})}{4\pi}          \nonumber \\
 & & \quad -2\epsilon Re[\psi^*(\Pi_{x}^2-\Pi_{y}^2)^2\psi]\bigg\rangle,
\end{eqnarray} 
where the brackets $\langle\rangle$ denote spatial averaging.

Using the generalizations of the Abrikosov identities derived with the help of 
Eq.~(\ref{virial}), the free energy can be cast into the compact form
\begin{equation}
\label{GLFE}
F=B^{2}-\frac{(\kappa-B)^{2}}{(2\kappa^{2}-1)\beta_{A}+1}\left\{1-\frac{\Gamma}{(2\kappa^{2}-1)
\beta_{A}+1}\right\},
\end{equation}
where $\beta_A$ is the usual Abrikosov parameter and 
$\Gamma$ is the correction to the isotropic result,
\begin{align}
\label{Gamma}
\begin{split}
\Gamma&=4\epsilon\bigg\{\frac{Re\langle|\psi|^{2}\psi^{*}(\Pi_{x}^{2}-\Pi_{y}^{2})^{2}\psi\rangle}
{\langle|\psi|^{2}\rangle^{2}}\\
&\phantom{=4\epsilon\bigg\{ }+\beta_{A}(2\kappa^{2}-1)\frac{Re\langle\psi^
{*}(\Pi_{x}^{2}-\Pi_{y}^{2})^{2}\psi\rangle}{\langle|\psi|^{2}\rangle}\bigg\}.
\end{split}
\end{align}
Throughout this paper we will use the convenient parameterization $\rho+i\sigma=(\overline{b}/\overline{a})
\exp({i\theta})$, where $\overline{b}$ and $\overline{a}$ are
the magnitudes of the two basis vectors of the vortex lattice and $\theta$ is the apex angle.

Equation (\ref{GL}) was solved perturbatively to \textit{second} order in the anisotropy parameter 
$\epsilon$, extending previous \textit{first} order results \cite{chang1,chang2,rosenstein}. 
The second order calculation is necessary to justify the use of perturbation theory
and at the same time to quantify the accuracy of the first order correction. 
More importantly, we uncover new and unexpected physics at second order:
a reorientation transition of the square vortex lattice with respect to the 
crystal lattice.  The quantities $\beta_{A}$ and $\Gamma$ are given in terms of 
rapidly converging sums involving $\rho$, $\sigma$, and $\phi$, 
\begin{align}
\begin{split}
\beta_{A}&=\beta_{0}+(\epsilon b)\beta_{1}+(\epsilon b)^{2}\left[\gamma_{1}+\gamma_{2}+5\beta_{1}-\frac{3}{4}
\beta_{0}\right],\\
\end{split}
\\
\begin{split}
\Gamma&=8\epsilon b^{2}\bigg\{\beta_{A}(2\kappa^{2}-1)[1+3(\epsilon b)]+\beta_{0}\left[1+\frac{3}{2}(\epsilon b)\right]\\
&\phantom{=8\epsilon b^{2}\bigg\{}+\beta_{1}[1+6(\epsilon b)]+2(\epsilon b)[\gamma_{1}+\gamma_{2}]\bigg\},
\end{split}
\end{align}
where the functions $\beta_{0}$, $\beta_{1}$, $\gamma_{1}$, $\gamma_{2}$ 
are periodic in $\rho$ with period $1$, and
are given by
\begin{align}
\beta_{0}&={\sum_{nm}}'A_{nm},\\
\beta_{1}&=Re\bigg\{e^{4i\phi}{\sum_{nm}}'A_{nm}\left[8\pi^{2}\sigma^{2}n^{4}-6\pi\sigma n^{2}+
\frac{3}{8}\right]\bigg\},\\
\begin{split}
\gamma_{1}&=Re\bigg\{e^{8i\phi}{\sum_{nm}}'A_{nm}\bigg[16\pi^{4}\sigma^{4}n^{8}-12\pi^{3}\sigma^{3}n^{6}\\
&\phantom{=Re\bigg\{e^{8i\phi}{\sum_{nm}}'}+\frac{3}{4}\pi^{2}\sigma^{2}n^{4}-\frac{45}{16}\pi\sigma n^{2}+
\frac{105}{256}\bigg]\bigg\},
\end{split}
\\
\begin{split}
\gamma_{2}&={\sum_{nm}}'A_{nm}\biggl[16\pi^{4}\sigma^{4}n^{4}m^{4}-12\pi^{3}\sigma^{3}n^{2}m^{2}(n^{2}+m^{2})
\notag
\end{split}
\\
\begin{split}
\phantom{\gamma_{2}}&\phantom{={\sum_{nm}}'A_{nm}\biggl[}+\frac{3}{4}\pi^{2}\sigma^{2}(n^{4}+m^{4}+36n^{2}m^{2})\\
\phantom{\gamma_{2}}&\phantom{={\sum_{nm}}'A_{nm}\biggl[}-\frac{45}{16}\pi\sigma(n^{2}+m^{2})
+\frac{105}{256}\biggr],
\end{split}
\end{align}
where $A_{nm}=\sqrt\sigma e^{2i\pi\rho(n^{2}-m^{2})}e^{-2\pi\sigma(n^{2}+m^{2})}$. The prime denotes a double
sum: one over $n$, $m$ and a second with $n$ and $m$ replaced by
$(n+1/2)$ and $(m+1/2)$ respectively.
We perform a numerical minimization of the free energy (\ref{GLFE})
with respect to $\rho$, $\sigma$, 
and $\phi$ to obtain the mean-field vortex lattice structure as a function 
of the field and temperature.  We find a continuous transition from a 
rhombic to square phase as $\epsilon$ is increased.
The second order correction only shifts the transition point by about
 1\% relative to the first order calculation, contrary to 
previous claims \cite{park} which neglected the lifting of the 
orientation degeneracy by the anisotropic term.   Our numerical investigations
show unambiguously that a consistent exploration of the relevant 
phase space and an accurate determination of the global minima of the 
GL free energy~(\ref{GLFE}) demands the inclusion of the 
the orientational degree of freedom $\phi$.
After the structural phase transition occurs the square vortex lattice 
is oriented along the 
[110] direction, and remains in this preferred orientation until a 
field $b_{r}$ is reached. At that point another continuous phase transition 
takes place, with the vortex lattice changing its orientation while 
retaining its square shape. 
At a given $b>b_{r}$, two equivalent configurations exist: 
one with orientation angle $\phi>\pi/4$, and one with $\phi'=\pi/2-\phi$. 
This reorientation transition occurs only at second order in the 
anisotropy $\epsilon$. 

Having determined the mean-field phase diagram of the system, 
we now turn to the elastic properties of the lattices.  Assuming 
that the displacement field 
$\mathbf{u}$  is small compared to typical lattice spacings, 
we calculate
the elastic moduli from the GL free energy (\ref{GLFE})
in the incompressible limit, $\nab\cdot{\bf u}=0$. 
For simplicity we have focused on straight and parallel vortices; 
we do not expect the structural transition to significantly affect the 
tilt modulus $c_{44}$. The relevant moduli are essentially non-dispersive 
\cite{brandt1}, allowing us to consider uniform deformations.
Now there are many ways to choose the elastic moduli from the 
independent components 
of the elastic tensor, and we will follow the lead of  
Miranovi\'{c} and Kogan \cite{mir}
and use as the four independent elastic moduli for a rhombic lattice the 
quantities $c_{sq}$, $c_{66x}$, $c_{66y}$, and $c_{r}$,
corresponding to four particular deformations. The so-called 
``squash" deformation is responsible for the transformation of the 
rhombic to a square vortex lattice, 
and  the corresponding squash modulus $c_{sq}$ vanishes at the 
transition point. The other relevant deformations
are the two pure shears in $x$ and $y$ directions 
(which determine $c_{66x}$, $c_{66y}$)
and the rotation of the lattice about the direction of the magnetic field
(which determines $c_{r}$).

All elastic moduli are obtained from the second derivative of the 
GL free energy (\ref{GLFE}).  The elastic energy can be written as
\begin{equation}
E=\frac{1}{2}\mu^{2}c(\mu)|_{\mu\rightarrow 0}=
\frac{1}{2}\mu^{2}\partial_{\mu}^{2}F|_{\mu\rightarrow  0},
\end{equation}
where $c(\mu)$ is the elastic modulus for each
particular deformation.  Each elastic modulus is obtained from
\begin{align}
\label{elm}
\begin{split}
c&=C_{\rho\rho}\rho_{\mu}^{2}+C_{\sigma\sigma}\sigma_{\mu}^{2}
+C_{\phi\phi}\phi_{\mu}^{2}\\
&+2C_{\rho\sigma}\rho_{\mu}\sigma_{\mu}+2C_{\rho\phi}\rho_{\mu}\phi_{\mu}
+2C_{\sigma\phi}\sigma_{\mu}\phi_{\mu},
\end{split}
\end{align}
where $C_{ij}$ are the terms originating from the derivatives of the 
GL free energy (\ref{GLFE})
\begin{multline}
\label{c}
C_{ij}=(2-\delta_{ij})\bigg\{\Phi(b^2-F){\beta_{A}}_{,ij}
+\Theta\bigg[\Gamma_{,ij}
-\Phi({\beta_{A}}_{,i}\Gamma_{,j}\\
+{\beta_{A}}_{,j}\Gamma_{,i}
+\Gamma{\beta_{A}}_{,ij}
-2\Phi\Gamma{\beta_{A}}_{,i}{\beta_{A}}_{,j})\bigg]\bigg\}.
\end{multline}
The derivatives in the right hand side of Eq.~(\ref{c}) are with respect to 
$i,j=\rho$, $\sigma$, $\phi$, 
$\Phi=(2\kappa^{2}-1)/[(2\kappa^{2}-1)\beta_{A}+1]$, and 
$\Theta=(1-b)^{2}/[(2\kappa^{2}-1)\beta_{A}+1]^{2}$.

\begin{figure}[ht]
\includegraphics[height=8.6cm,width=7.4cm,angle=-90,clip]{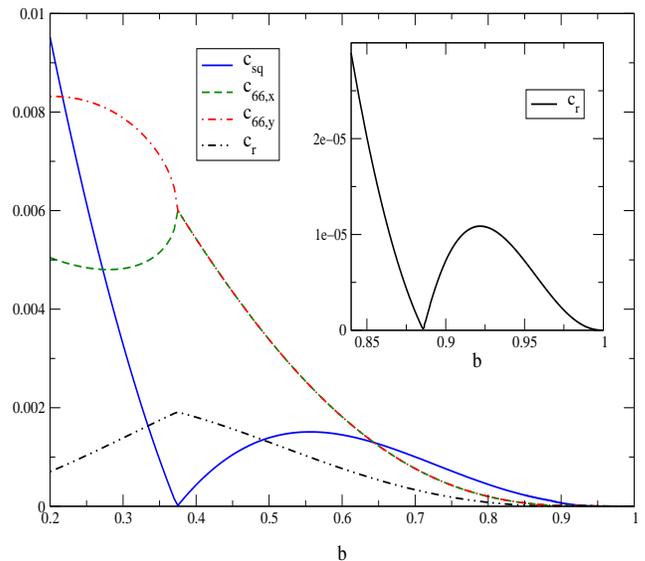}
\caption{The elastic moduli versus the reduced field $b$ for
$\epsilon=0.11$. In the inset we show the region
where the reorientation transition happens, as is signaled by the
vanishing of the rotation modulus $c_{r}$.}
\label{moduli}
\end{figure}

The calculated elastic moduli for $\epsilon=0.11$ and $\kappa=10$
(appropriate for the borocarbides) are shown in 
Fig.~\ref{moduli}.  The transition from a rhombic to a square vortex 
lattice is signaled by the vanishing of the
squash modulus.  At the same point the two
shear moduli merge into one because of the higher symmetry of the 
square phase. At a still higher field we
observe the reorientation transition, which occurs when $c_{r}=0$. This
instability of the vortex lattice with respect to rotations also 
exists in the nonlocal London model \cite{mir}. 
There are experimental indications for this 
instability, although its investigation has been 
hampered so far by the high fields required for its 
observation \cite{vinnikov}. 
This reorientation transition is sensitive to the value of $\kappa$, 
and from our results we predict that this transition will not occur 
in the cuprates ($\kappa\sim100$), as the transition point where 
$c_{r}=0$ moves above $H_{c2}$.

\begin{figure}[ht]
\includegraphics[height=8.6cm,width=7.4cm,angle=-90,clip]{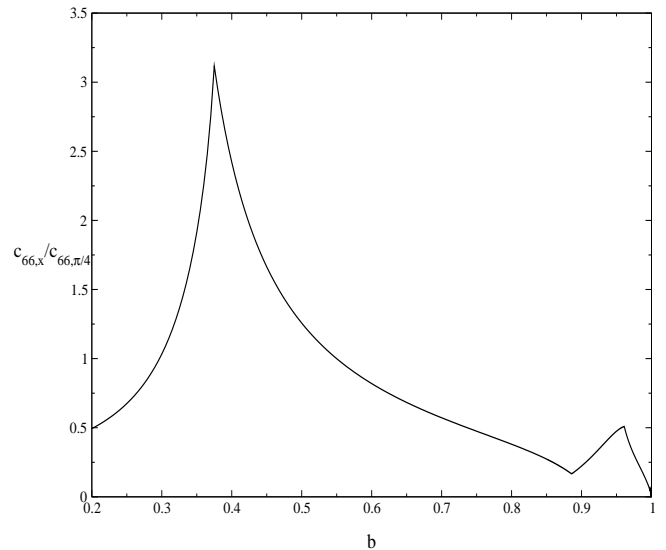}
\caption{The ratio $c_{66x}/c_{66,\frac{\pi}{4}}$
versus the reduced field $b$.}
\label{ratio}
\end{figure}

We also note that anisotropic elasticity with a softer shear modulus 
for shearing along the sides of the square lattice than along the 
diagonals emerges naturally in this model. This
response of the vortex lattice is measured by the ratio 
$c_{66x}/c_{66,\frac{\pi}{4}}$, which has a
maximum at the transition as can be seen in Fig.~\ref{ratio}. 
It has been suggested \cite{vinnikov}
that this behavior explains the anisotropic orientational long range 
order observed in decoration patterns in LuNi$_{2}$B$_{2}$C, which 
manifests itself as a significant difference in the correlation
lengths along the $[110]$ and $[\bar{1}10]$ directions.

\begin{figure}[ht]
\includegraphics[height=8.6cm,width=8.4cm,angle=-90,clip]{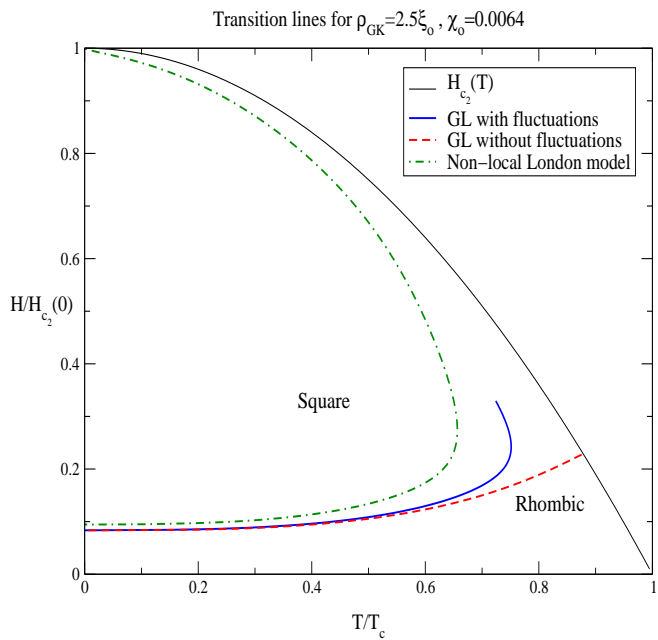}
\caption{The phase boundaries in the $H$-$T$ plane separating the 
square and rhombic vortex lattice obtained
from GL theory and the nonlocal London model. The
line intercepting $H_{c2}(t)$ is the mean field GL result.}
\label{lines}
\end{figure}

We next turn to the effects of thermal fluctuations on the vortex
lattice structure.  Following Gurevich and Kogan \cite{gurevich}, 
we will make the simplifying assumption that harmonic fluctuations 
renormalize the three variables $\rho$, $\sigma$, 
and $\phi$, which characterize the structure of the vortex 
lattice, with the result that   
\begin{align}
\label{rn}
\overline\rho&=\rho[1-\sigma(2-\rho)\overline{u^{2}}],\\
\label{sn}
\overline\sigma&=\sigma[1-\sigma(2-\rho)\overline{u^{2}}],\\
\label{fn}
\overline\phi&=\phi+\sigma\cot{\phi}\;\overline{u^{2}},
\end{align}
where overbars denote thermal averages. The rhombic to
square transition is then determined as the locus of points in
the $H$--$T$ plane where $c_{sq}=0$. For that purpose, the ``squash''
elastic modulus $c_{sq}$ is recalculated from Eq.~(\ref{elm}) using the renormalized variables
$\overline\rho$, $\overline\sigma$, $\overline\phi$.
 This is not performed self-consistently, as we do not take
into account the effects of the fluctuations on the 
elastic moduli themselves.
This approach is justified due to the fact
that the mean squared displacement of the vortex $\overline{u^{2}}$ is
finite at $H_{\square}(T)$ \cite{gurevich}.
 There are two competing parameters, the strength of the thermal 
fluctuations $\chi$ and the anisotropy parameter
$\epsilon$ which is related to the nonlocality 
parameter $\rho_{GK}$ of the nonlocal London
model as $\epsilon=\left(\rho_{GK}/\xi_0\right)^{2}
(1-t^{2})/12$.  To compare with the nonlocal London model results, 
we take \cite{gurevich} 
$\lambda(T)=\lambda_0/\sqrt{1-t^{2}}$ and 
$\xi(T)=\xi_0/\sqrt{1-t^{2}}$,
where $t=T/T_{c}$, so that $\chi=\chi_{0}t/\sqrt{1-t^{2}}$. 
The dimensionless mean squared displacement is then 
\begin{equation}
\overline{u^{2}}=\frac{\chi_{0}\eta}{\sqrt{(1-t^2)}
}\frac{bt}{[(1-b)^{3}\ln(1+\frac{1}{\sqrt{2b}})]^{\frac{1}{2}}},
\end{equation}
with $\eta\sim 3$ for LuNi$_{2}$B$_{2}$C \cite{gurevich}. 
The results of the numerical solution are shown in
Fig.~\ref{lines}, for the same parameter values used 
by Gurevich and Kogan \cite{gurevich}, namely, 
$\rho_{GK}=2.5\xi_{0}$ and $\chi_{0}=0.0064$. 
With fluctuations, the phase boundary curves backward and displays
the ``reentrant'' behavior that was also obtained by Gurevich and 
Kogan. The physical explanation of this feature is simple: thermal 
fluctuations tend to wash out effects of the anisotropy, so that
the preferred symmetry of the lattice is rhombic. 
Note that the abrupt termination of the transition line for the 
GL theory is an artifact of the perturbative treatment of the 
anisotropy; at that point the correction term $\Gamma$ ceases to be small and 
the extended GL theory becomes unstable. While this 
shortcoming of the model could be
remedied by including higher order terms in the GL free energy, 
the overall qualitative features of the phase diagram 
are expected to remain unchanged. One such choice would be
an additional isotropic term. Nevertheless, the complications 
introduced in the analysis by the unnecessary extra parameter,
 would clearly outweigh any potential benefits.

We have used the GL theory with an anisotropic, symmetry breaking term 
to study the evolution of the vortex lattice and particularly the structural 
phase transition and its connection with
the elastic response of the vortex lattice. We have shown that 
to obtain the correct phase diagram it is necessary to calculate 
properties up to second order in the anisotropy;  at this order
a rotational instability of the square vortex lattice  appears. 
We predict that this instability should be observable in the borocarbides
but not in the cuprate superconductors. 
Including thermal fluctuations in the model we were able to show that the 
rhombic to square transition line exhibits reentrant behavior, similar
to what is obtained within the nonlocal London model, and in qualitative
agreement with SANS studies of LuNi$_2$B$_2$C. 

We would like to thank V. G. Kogan, A. Gurevich and P. J. Hirschfeld 
for useful discussions.  This work was supported by NSF DMR-9978547.

\end{document}